\documentclass[12pt]{iopart}
\pdfoutput=1

\usepackage{url}
\usepackage{graphicx}
\usepackage{color}
\usepackage{latexsym}
\usepackage{enumerate}
\usepackage{multirow}
\usepackage{twoopt}
\usepackage{array}
\usepackage{alltt}

\definecolor{orange}{rgb}{0.8,0.2,0.2}

\begin{document}

\title[Network Strategies in Election Campaigns]{Network Strategies in Election Campaigns}

\author{Marco Alberto Javarone}

\address{Dept. of Mathematics and Computer Science, University of Cagliari, 09123 Cagliari, Italy \\
DUMAS - Dept. of Human and Social Sciences, University of Sassari, 07100 Sassari, Italy}

\begin{abstract}
This study considers a simple variation of the voter model with two competing parties. In particular, we represent the case of political elections, where people can choose to support one of the two candidates or to remain neutral. People operate within a social network and their opinions depend on those of the people with whom they interact. Therefore, they may change their opinions over time, which may mean supporting one particular candidate or none. Candidates attempt to gain people's support by interacting with them, whether they are in the same social circle (i.e. neighbors) or not. In particular, candidates follow a strategy of interacting for a time with people they do not know (that is, people who are not their neighbors). Our analysis of the proposed model sought to establish which network strategies are the most effective for candidates to gain popular support. We found that the most suitable strategy depends on the topology of the social network. Finally, we investigated the role of charisma in these dynamics. Charisma is relevant in several social contexts, since charismatic people usually exercise a strong influence over others. Our results showed that candidates' charisma is an important contributory factor to a successful network strategy in election campaigns.
\end{abstract}

\pacs{89.75.-k, 89.65.-s, 89.75.Fb}
\maketitle
\section{Introduction}\label{intro}
In the last years, opinion dynamics~\cite{loreto01} has attracted the attention of many scientists and several models, to study the formation and the spreading of opinions, have been developed (e.g., ~\cite{galam01,galam02,krapviski01,sznajd01,holme01,bianconi01}). 
In these dynamics, interactions among individuals and the topology of their network play a fundamental role~\cite{loreto01,miguel01,barrat01}. 
One of the most simple models of opinion dynamics is the voter model~\cite{redner01,galam02,galam03,serrano01}. The latter describes a set of agents that change opinion over time by interacting among themselves.
The voter model allows to represent the evolution of a population toward consensus in the presence of different opinions. 
In general, from a physical perspective, by this model it is possible to identify phase transitions in the system, as from a disordered state to an ordered one~\cite{mobilia01,vespignani01}; although as shown in~\cite{schweitzer01} also non-linear dynamics, that entails the system reaches a final phase characterized by the coexistence of different opinions, can be introduced.
Moreover, the voter model can be implemented in several ways, with the aim to catch a particular character or behavior of real systems, as political elections~\cite{gross01,miguel02,petersen01} and, more in general, competitions~\cite{galam04,clifford01,miguel03}.
In this work, we introduce a variant of the classical voter model, putting our attention on the case of political elections, in order to study the best strategies to gain the popular consensus.
In the proposed model, there are two competitors (or candidates) that try to convince a community of agents. In turn, agents are neutral or they have a preference for one competitor. Therefore, we consider a system with three possible opinions~\cite{marvel01}.
Agents are arranged in a network and they change opinion over time, by considering those of their neighbors.
During the evolution of the system, competitors try to affect the opinion of agents by defining temporal connections with them. In particular, agents temporarily connected with competitors consider them as normal neighbors while compute their next opinion.
Therefore, each competitor considers very important to identify best agents for generating these temporal connections. In this context, best agents are those that allow to increase the candidate's consensus as fast as possible in the whole population. 
It is worth to note that the described dynamics is based on the structure of underlying adaptive networks~\cite{gross02,gross03,zschaler01}.
We perform a comparison among different network strategies, used to perform the selection of agents.
Results of numerical simulations show a relation between the best strategy and the topology of the agent network. Moreover, we investigate whether the definition of network strategies should consider also the competitors' charisma, as this quality is considered fundamental in social contexts. 
The remainder of the paper is organized as follows: Section~\ref{sec:model} introduces the proposed model, for studying network strategies to gain the popular consensus. Section~\ref{sec:results} shows results of numerical simulations. Finally, Section~\ref{sec:conclusions} ends the paper.
\section{The  Model} \label{sec:model}
\begin{figure}
\centering
\includegraphics[width=3.0in]{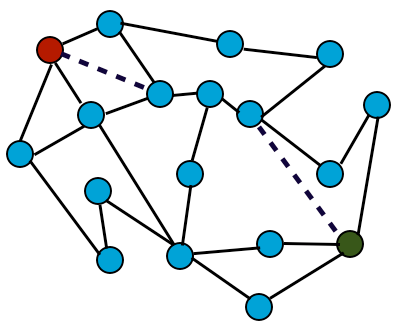}
\caption{\label{fig:figure_1} Two competitors (i.e., red and green nodes) generate a temporal edge, indicated by a dotted line, following a strategy: the red node uses the strategy \textbf{S2} (i.e., it selects $2$nd degree connections), whereas the green node uses the strategy \textbf{S3} (i.e., it selects $3$rd degree connections).}
\end{figure}

We introduce a simple variant of a voter model with two competitors, e.g., two candidates during an electoral campaign. Competitors aim to gain the popular consensus in a population of agents. In turn, agents are arranged in a network and they can interact with their neighbors. Moreover, each agent has an opinion, i.e., it can be neutral or to have a preference for one competitor.  
Opinions are mapped to states, hence agents in the state $0$ are neutral, whereas agents in the state $1$ and $2$ have a preference for the competitor $1$ and $2$, respectively. 
Although three-state voter models have been already analyzed by other authors (see~\cite{szolnoki01}), usually classical implementations consider only two opinions, i.e., the two-state variant is much more studied.
Agents change opinion over time by considering those of their neighbors (obviously, competitors never change opinion). 
At time $t=0$, all agents are in a neutral state (i.e., $0$), with the exception of the two competitors that are in the state $1$ and $2$, respectively.
Then, at each time step, agents change their state (i.e., opinion) according to the following transition probabilities:
\begin{equation} \label{eq:transition_probabilities_zero}
\cases{
p_{x\to y} = \sigma_y \\ 
 p_{x} = 1 - \sum_{i=0 | i \neq x}^{2}\sigma_i
}
\end{equation}
\noindent with $p_x \to y$ transition probability to change from the $x$th state to the $y$th state and  $p_x$ probability to remain in the same state. 
The value of $\sigma_y$ is computed as $\sigma_y = n_y/n_t$, with $n_y$ number of neighbors in the state $y$th and $n_t$ total number of neighbors (i.e., the degree of the agent). Eventually, the summation to compute $p_x$ considers the densities $\sigma_i$ of neighbors having all the feasible states different from the $x$th state.    
In so doing, at each time step, the agents' states are defined by using a weighted random selection with the transition probabilities (Eq.~\ref{eq:transition_probabilities_zero}) used as weights.
Therefore, the evolution of the system is described by the following equations:

\begin{equation} \label{eq:evolution_zero}
\cases{
N_{0}(t+1) =  N\cdot\left \{ \sum_{i=1|o_{i}\neq0}^{N} \sigma_{0}^{i}(t) - \left [ \sum_{i=1| o_{i}=0}^{N} \sigma_{1}^{i}(t) + \sigma_{2}^{i}(t) \right ] \right \} + N_{0}(t) \\ 
N_{1}(t+1) =  N\cdot\left \{ \sum_{i=1|o_{i}\neq1}^{N} \sigma_{1}^{i}(t) - \left [ \sum_{i=1|o_{i}=1}^{N} \sigma_{0}^{i}(t) + \sigma_{2}^{i}(t) \right ] \right \} + N_{1}(t)\\
N_{2}(t+1) =  N\cdot\left \{ \sum_{i=1|o_{i}\neq2}^{N} \sigma_{2}^{i}(t) - \left [ \sum_{i=1|o_{i}=2}^{N} \sigma_{0}^{i}(t) + \sigma_{1}^{i}(t) \right ] \right \} + N_{2}(t)
} 
\end{equation}
\noindent with $N_{x}$ number of agents in the $x$th state and $\sum_{i=1|o_{i}=x}$ that indicates the $i$th agent having a state $o_i$ equal (or different) to $x$. 
Competitors do not change their state and they try to gain the consensus of the population. In particular, they generate temporal connections with agents that are not their neighbors, with the aim to affect the value of their transition probabilities. 
These temporal connections last only for one time step and each competitor generates, every time, a number of temporal connections equal to its degree (i.e., the number of its neighbors). Therefore, agents temporarily connected with a competitor compute their transition probabilities as follows:
\begin{equation} \label{eq:transition_probabilities_one}
p_{x\to y} = \sigma_{y}^{t} 
\end{equation}
\begin{equation} \label{eq:stay_one}
p_x = 1 - p_{x \to y}
\end{equation}
\noindent with $\sigma_{y}^{t}$ temporal density of neighbors in the $y$th state (i.e., the state of the competitor that contacted the agent), computed as:
\begin{equation} \label{eq:temporal_density}
\sigma_{y}^{t} = \frac{n_y+1}{n_t + 1} 
\end{equation}
In so doing, the equations to describe the evolution of the system become:

\begin{equation} \label{eq:evolution_one}
\cases{ 
N_{1}(t+1)^{T} =  N_{1}(t+1) + k_{1}\cdot\left \{ \sum_{j=1}^{k_{1}} \sigma_{1}^{t A_1\left[j\right]}(t)\right \} - k_{2}\cdot\left \{ \sum_{j=1|o_{j}=1}^{k_{2}} \sigma_{2}^{t A_2\left[j\right]}(t)\right \} \\
N_{2}(t+1)^{T} =  N_{2}(t+1) + k_{2}\cdot\left \{ \sum_{j=1}^{k_{2}} \sigma_{2}^{t A_2\left[j\right]}(t)\right \} - k_{1}\cdot\left \{ \sum_{j=1|o_{j}=2}^{k_{1}} \sigma_{1}^{t A_1\left[j\right]}(t) \right \}\\
N_{0}(t+1)^{T} = N - N_{1}(t+1)^{T} - N_{2}(t+1)^{T}
}
\end{equation}
\noindent with $N_{x}(t+1)^{T}$ number of agents in the $x$th state, considering the temporal connections, and $k_1$, $k_2$ degree of the competitor $1$ and $2$, respectively. 
The exponent of $\sigma_x^{t}$ in Eq.~\ref{eq:evolution_one}, i.e., $A_x\left[j\right]$, represents the $j$th agent among those selected by the $x$th competitor, for generating temporal connections at time $t$.
During the electoral campaign, at each time step, competitors have to select the most useful agents to generate temporal connections. In order to perform this selection, competitors can use one of the following network strategies:
\begin{itemize}
\item \textbf{S0}. Random selection;
\item \textbf{S1}. Random weighted selections, using the degree of agents as weights;
\item \textbf{S2}. $2$nd degree connections: agents at distance $2$ (i.e., neighbors of their neighbors); 
\item \textbf{S3}. $3$rd degree connections: agents at distance $3$.
\end{itemize}
Figure~\ref{fig:figure_1} shows an example where two competitors generate a temporal connection by using the strategy \textbf{S2} and the strategy \textbf{S3}, respectively.
Strategies \textbf{S0} and \textbf{S1} can be defined as ``global strategies'', as competitors consider the whole network to select agents. Moreover, by using the strategy \textbf{S1}, agents with high degree have a higher probability to be selected. 
On the other hand, strategies \textbf{S2} and \textbf{S3} can be defined as ``local strategies'', as competitors select agents by considering only the small portion of the network around them (i.e., friends of friends, and so on).
In order to evaluate whether a best network strategy can be identified, among those listed above, we analyze the proposed model by using scale-free networks and small-world networks to connect the agents.
\section{Results} \label{sec:results}
\begin{figure*}[!ht]
\centering
\includegraphics[width=6.0in]{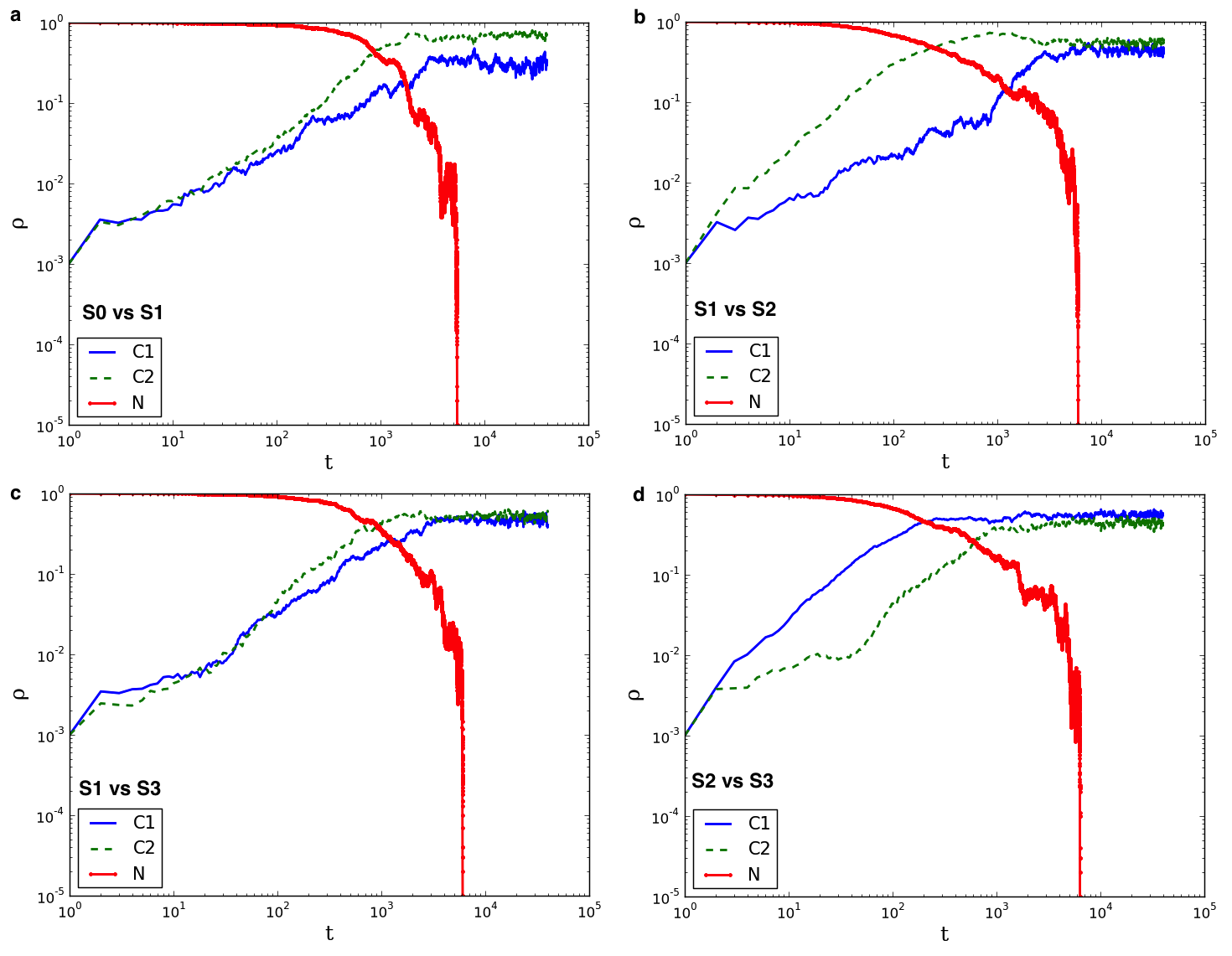}
\caption{\label{fig:figure_2} Comparison among the density of agents in the state $0$ (labeled as $N$ in the legend, i.e., neutral agents), in the state $1$ (labeled as $C_1$ in the legend, i.e., Competitor $1$) and in the state $2$ (labeled as $C_2$ in the legend) over time, performed on scale-free networks. Results are averaged over $20$ different realizations. \textbf{a}) Competitor $1$ uses the strategy \textbf{S0} and competitor $2$ uses the strategy \textbf{S1}. \textbf{b}) Competitor $1$ uses the strategy \textbf{S1} and competitor $2$ uses the strategy \textbf{S2}. \textbf{c}) Competitor $1$ uses the strategy \textbf{S1} and competitor $2$ uses the strategy \textbf{S3}. \textbf{d}) Competitor $1$ uses the strategy \textbf{S2} and competitor $2$ uses the strategy \textbf{S3}.}
\end{figure*}

\begin{figure*}[!ht]
\centering
\includegraphics[width=6.0in]{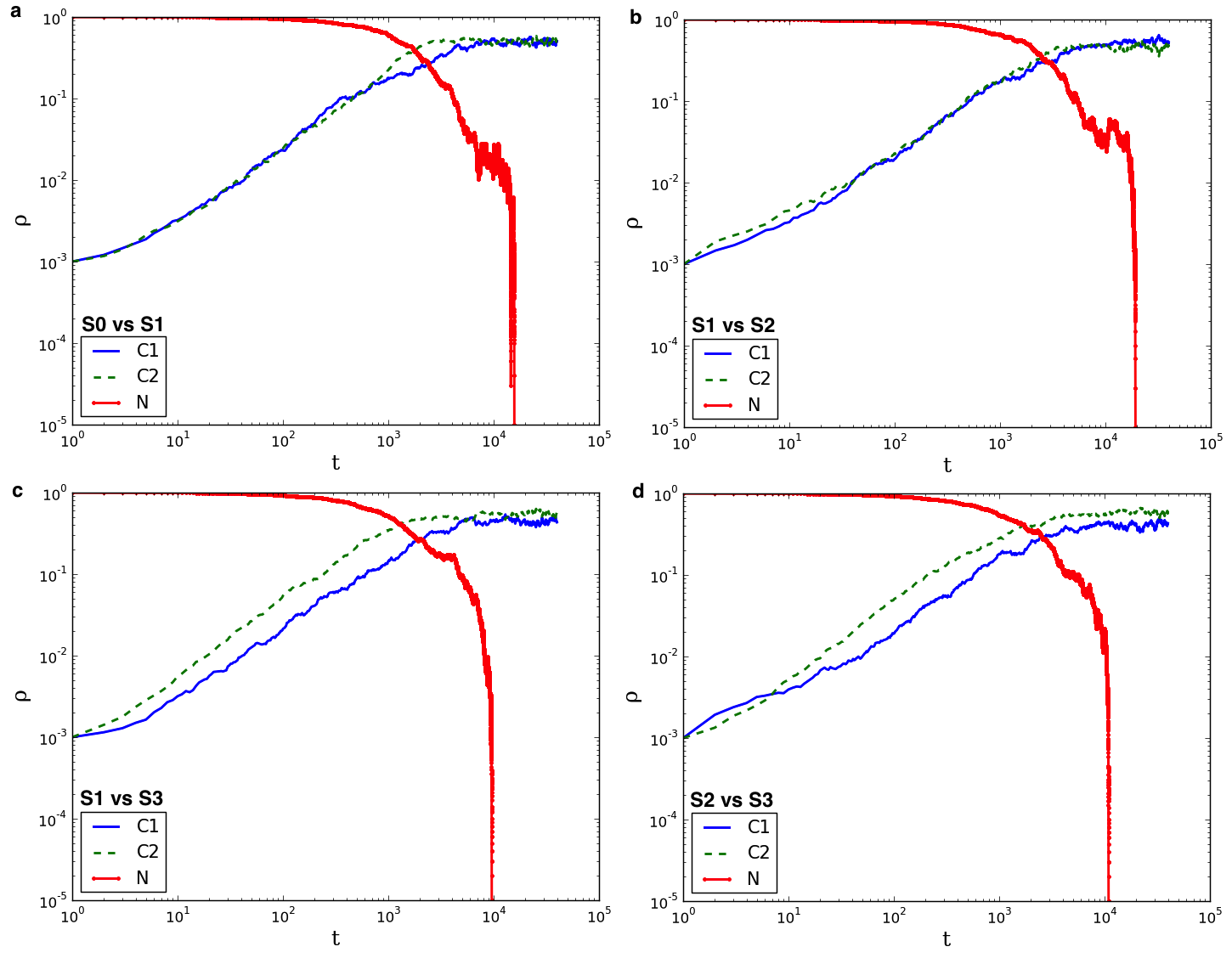}
\caption{\label{fig:figure_3} Comparison among the density of agents in the state $0$ (labeled as $N$ in the legend, i.e., neutral agents), in the state $1$ (labeled as $C_1$ in the legend, i.e., Competitor $1$) and in the state $2$ (labeled as $C_2$ in the legend) over time, performed on small-world networks. Results are averaged over $20$ different realizations. \textbf{a}) Competitor $1$ uses the strategy \textbf{S0} and competitor $2$ uses the strategy \textbf{S1}. \textbf{b}) Competitor $1$ uses the strategy \textbf{S1} and competitor $2$ uses the strategy \textbf{S2}. \textbf{c}) Competitor $1$ uses the strategy \textbf{S1} and competitor $2$ uses the strategy \textbf{S3}. \textbf{d}) Competitor $1$ uses the strategy \textbf{S2} and competitor $2$ uses the strategy \textbf{S3}.}
\end{figure*}

We performed many numerical simulations of the proposed model with the aim to identify the best network strategy to gain the popular consensus. Agents have been arranged in scale-free networks, generated by the Barabasi-Albert model (BA model hereinafter)~\cite{barabasi01}, and in small-world networks, generated by the Watts-Strogatz model (WS hereinafter)~\cite{watts01}. In particular, to achieve small-world networks, we start from a $2$-dimensional regular lattice with $6$ neighbors per node, then we rewire with probability $\beta = 0.1$ each edge at random. 
Finally, both kinds of network (i.e., scale-free and small-world) have a number of agents $N=10^4$, provided with an average degree $\langle k\rangle = 6$. 
Moreover, we performed further simulations in scale-free networks with $N >\ 10^4$, to observe the effects caused by the presence of a greater number of hubs --see Appendix.
We recall that scale-free networks, generated by the BA model, have a degree distribution $P(k)$ characterized by a scaling parameter $\gamma \approx 3$.
In order to compare network strategies, we consider the number of agents that have a preference for each competitor and the number of neutral agents. In particular, we analyze the variation of the density $\rho$ of agents, in these three states, over time.
In Figures~\ref{fig:figure_2} and~\ref{fig:figure_3}, we report the comparison among four different simulations, performed on scale-free networks and small-world networks, respectively.
A first information, achieved analyzing the curves $(\rho,t)$, representing agents in different states, is that the number of neutral agents falls to zero after about $5.5\cdot10^3$ time steps in scale-free networks and after about $1.2\cdot10^4$ time steps in small-world networks. After that, in both kinds of network, the system seems to reach a steady-state, characterized by small fluctuations of densities between the two states $1$ and $2$.
Moreover, in the curves $(\rho,t)$, we identify two important points, called $T_1$ and $T_2$. These points constitute the intersections between the density of neutral agents and those of agents in the other states. The point $T_1$ is the intersection between neutral agents and agents with the preference for the competitor $1$, whereas $T_2$ is the intersection between neutral agents and agents with the preference for the other competitor (i.e., the $2$).
As discussed below, points $T_1$ and $T_2$ are useful to compare the network strategies.
\subsection{Comparison among network strategies}
\begin{figure*}[ht]
\centering
\includegraphics[width=6.0in]{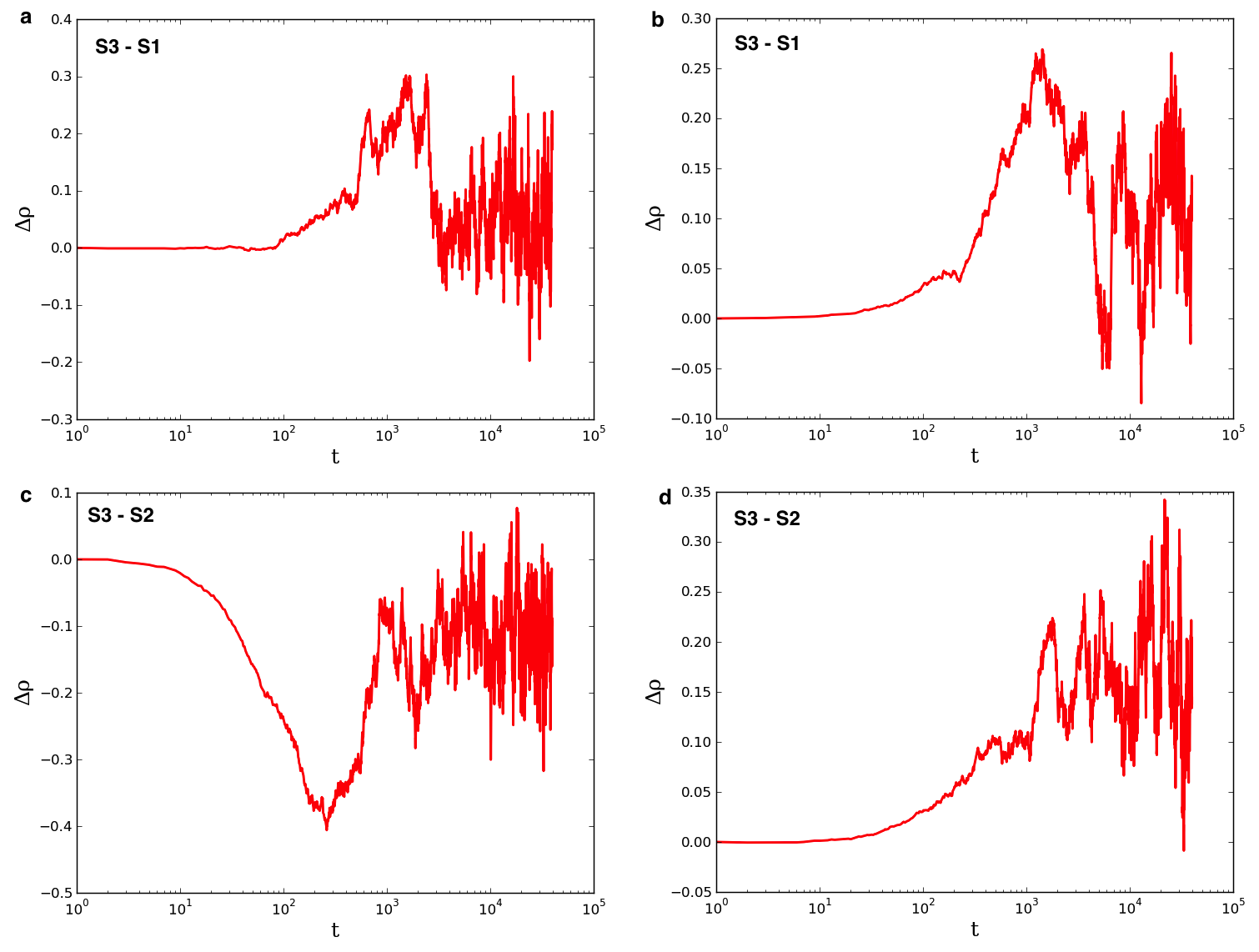}
\caption{\label{fig:figure_4} Difference between densities of agents $\Delta\rho$ in the state $1$ and $2$, varying the strategies adopted by the two competitors. Results are averaged over $20$ different realizations. \textbf{a}) Results achieved in scale-free networks, when the competitors use strategies \textbf{S3} and \textbf{S1}, respectively. \textbf{b}) Results achieved in small-world networks, when the competitors use the strategies \textbf{S3} and \textbf{S1}, respectively. \textbf{c}) Results achieved in scale-free networks, when the competitors use the strategies \textbf{S3} and \textbf{S2}, respectively. \textbf{d}) Results achieved in small-world networks, when the competitors use the strategies \textbf{S3} and \textbf{S2}, respectively.}
\end{figure*}

\begin{figure*}[ht]
\centering
\includegraphics[width=6.0in]{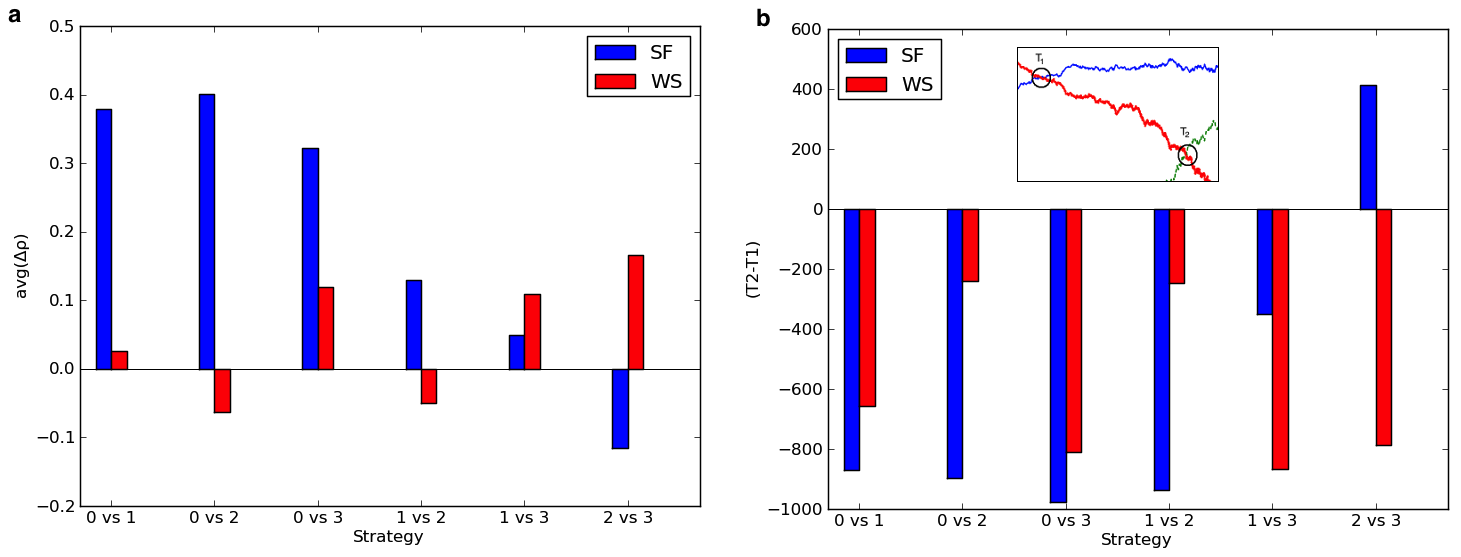}
\caption{\label{fig:figure_5} Comparison between results achieved in scale-free networks (blue bars) and those achieved in small-world networks (red bars), varying the network strategy. Results are averaged over $20$ different realizations. \textbf{a}) $avg(\Delta\rho)$, i.e., average difference between density of agents in the two states, $1$ and $2$. \textbf{b}) Difference between points $T_2$ and $T_1$, indicated in the inset (that shows an enlargement of a diagram $(\rho,t)$).}
\end{figure*}

A useful parameter, to compare network strategies, is the difference of densities $\Delta\rho$ between agents in the state $1$ and agents in the state $2$, over time --see Figure~\ref{fig:figure_4}. 
The topology of the agents network seems to play a crucial role, as we observe by comparing results shown in panels \textbf{c} and \textbf{d} of Figure~\ref{fig:figure_4}, related to scale-free networks and small-world networks, respectively. 
In particular, the strategy \textbf{S2} is better than the strategy \textbf{S3} in scale-free networks, but just the opposite occurs in small-world networks (i.e., the strategy \textbf{S3} is the best one).
Therefore, we computed the average value of $\delta\rho$, comparing all strategies in both kinds of networks --see panel \textbf{a} of Figure~\ref{fig:figure_5}. As discussed before, the points $T_1$ and $T_2$ of diagrams $(\rho,t$) can provide an information about the speed of competitors in the earning of the global consensus. In particular, as shown in panel \textbf{b} of Figure~\ref{fig:figure_5}, we computed the difference $T_2 - T_1$ for each curve $(\rho,t)$.
Values of $avg(\Delta\rho)$ highlight that, in scale-free networks, local strategies are better than global ones. In particular, the best strategy is \textbf{S2}. Instead, considering the global strategies, the \textbf{S1} is much more better than \textbf{S0}. 
On the other hand, in small world networks, we found that the best strategy is \textbf{S3}, followed by the strategy \textbf{S1}. Therefore, also in this case a local strategy is more efficient than global ones. It is interesting to note that the strategy \textbf{S2} yields optimal results in scale-free networks, but it is the worst strategy (among those analyzed) in small-world networks. Eventually, comparing the global strategies, we found that \textbf{S1} is always better than \textbf{S0}, in particular in scale-free networks, due to the presence of hubs (i.e., nodes with a high degree).
A further information is provided by the histogram $(T_2 - T_1)$ in panel \textbf{b} of Figure~\ref{fig:figure_5}. In particular, we can evaluate which are the faster strategies to gain the popular consensus. 
As discussed before, after that the number of neutral agents falls to zero, the system reaches almost a steady-state, with small differences between the density of agents in states $1$ and $2$. Therefore, as the time is an important variable in competitions as political elections~\cite{hillygus01}, a good strategy allows also to obtain the consensus in a few time steps. 
As result of this analysis, we found that best strategies, identified in the histogram $avg(\Delta\rho)$, are also faster than the other ones. Furthermore, it is worth to highlight that, although \textbf{S2} is weaker than global strategies in small-world networks (according to values of $avg(\Delta\rho)$), it yields a fast increasing of global consensus. Hence, in the event the time variable is critical (i.e., competitors have a few time to gain the consensus), local strategies are better than global ones.
\subsection{Charismatic Competitors}
\begin{figure*}
\centering
\includegraphics[width=6.0in]{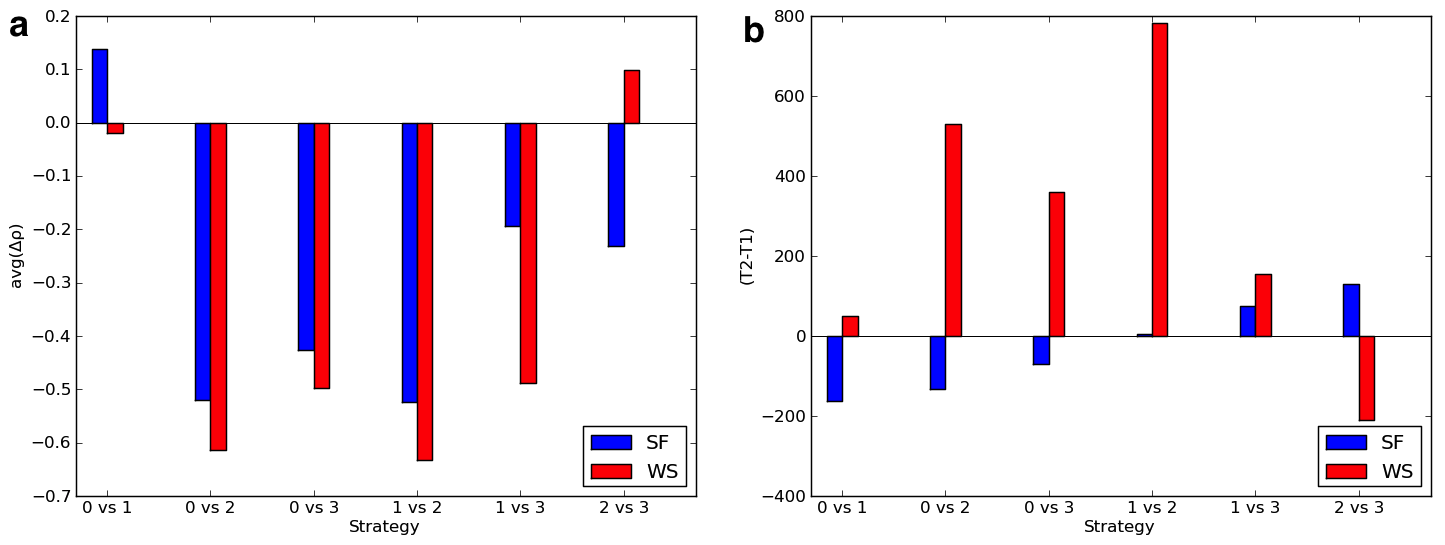}
\caption{\label{fig:figure_6} Comparison between results achieved in scale-free networks (blue bars) and those achieved in small-world networks (red bars), varying the network strategy and considering charismatic competitors. Results are averaged over $20$ different realizations. \textbf{a}) $avg(\Delta\rho)$, i.e., average difference between density of agents in the two states, $1$ and $2$. \textbf{b}) Difference between points $T_2$ and $T_1$.}
\end{figure*}

According to recent studies~\cite{merolla01,keyman01}, the politicians' charisma plays an important role in the achievement of the popular consensus. In general, the charisma is a quality deemed relevant in several social contexts as charismatic people are able to exercise a strong influence on other people.
Therefore, here we investigate the proposed model considering charismatic competitors. 
In particular, we modify the transition probabilities of temporarily connected agents as follows:
\begin{equation} \label{eq:transition_probabilities_charisma}
p_{x\to y} = 
\cases{1  & \mbox{if $x = 0$ } \\ 
\frac{1}{2}  & \mbox{if $x \neq 0$ }
}
\end{equation}
\noindent whereas, $p_x$, i.e., the probability that the temporarily connected agents remain in the same state, is always computed by Eq.~\ref{eq:stay_one}.
In so doing, a charismatic competitor gains always the consensus of neutral agents, whereas it has the $50\%$ of probabilities to gain the consensus of agents that prefer its opponent.
Figure~\ref{fig:figure_6} shows results achieved in both kinds of network (i.e., scale-free and small-world) varying the network strategies played by competitors.
It is interesting to note that the presence of charismatic competitors strongly affects results. In particular, considering the histogram $(\Delta\rho)$ (panel \textbf{a} of Figure~\ref{fig:figure_6}), global strategies are better than local ones in both kinds of network. In scale-free networks the best strategy is \textbf{S1}, whereas in small-world networks the best one seems to be \textbf{S0}. 
Notwithstanding, observing the histogram $(T_2-T_1)$, we can see that in scale-free networks there are small temporal differences between strategies. Therefore, from this point of view, all strategies are similar. Instead, in small-world networks we found that best strategies are also the fastest ones.
Finally, even if we consider the presence of charismatic competitors, the topology of networks still affects results. 
\section{Discussion and Conclusions} \label{sec:conclusions}
In this work, we analyze network strategies to gain the popular consensus in the presence of two competitors.
We define a simple variation of the voter model with agents that change opinion according to transition probabilities, computed considering the opinions of their neighbors. Moreover, we let competitors interact temporarily also with agents that are not their neighbors, with the aim to affect their opinion. Therefore, as observed before, the proposed model is based on an adaptive network.
In particular, at each time step, competitors select a number of agents, equal to their degree, to generate temporal connections. This selection is performed by using a network strategy. Competitors can choose between global strategies, i.e., random selection and weighted random selection (to select agents with a high degree), and local strategies, i.e., their $2$nd connections degree and $3$rd connections degree.
Simulations have been performed by arranging agents in scale-free networks and in small-world networks.
Results highlight that the topology of networks strongly affects the outcomes of the model. In particular, in scale-free networks the best strategy to select agents is \textbf{S2}, i.e., $2$nd connections degree. On the other hand, in small-world networks is more efficient the strategy \textbf{S3}, i.e., $3$rd connections degree. 
In general, we found that local strategies are more advantageous than global ones in both kinds of network. In particular, although the strategy \textbf{S2} seems unfavourable in small-world networks, it is faster than both global strategies. We recall that the term ``fast'', in this context, is used to identify strategies that allow to increase the global consensus in a few time steps.
Furthermore, we performed simulations considering ``charismatic'' competitors. We model the charisma of competitors by using their probability to convince temporarily connected agents. In particular, this probability is $1$ in the event they interact with neutral agents, whereas it is equal to $0.5$ in the event they interact with agents that have a preference for their opponent.
In so doing, we found that global strategies are better than local ones. In small-world networks both histograms, i.e., $\Delta\rho$ and $T_2-T_1$, show that \textbf{S0} and \textbf{S1} are better than local strategies, and moreover, they yield similar results. 
On the other hand, in scale-free networks, global strategies are still better than local ones but, from a temporal perspective, there are small differences, i.e., all strategies allow to convince many agents in a similar number of time steps.
In order to conclude, results highlight that both the topology of the agent network and the charisma of competitors should be considered to plan a successful strategy during electoral campaigns. 

\section*{Acknowledgments}
The author wishes to thank Ginestra Bianconi for her helpful comments and suggestions. This work was supported by Fondazione Banco di Sardegna.

\section*{Appendix}\label{sec:appendix}
\begin{figure*}[!ht]
\centering
\includegraphics[width=6.0in]{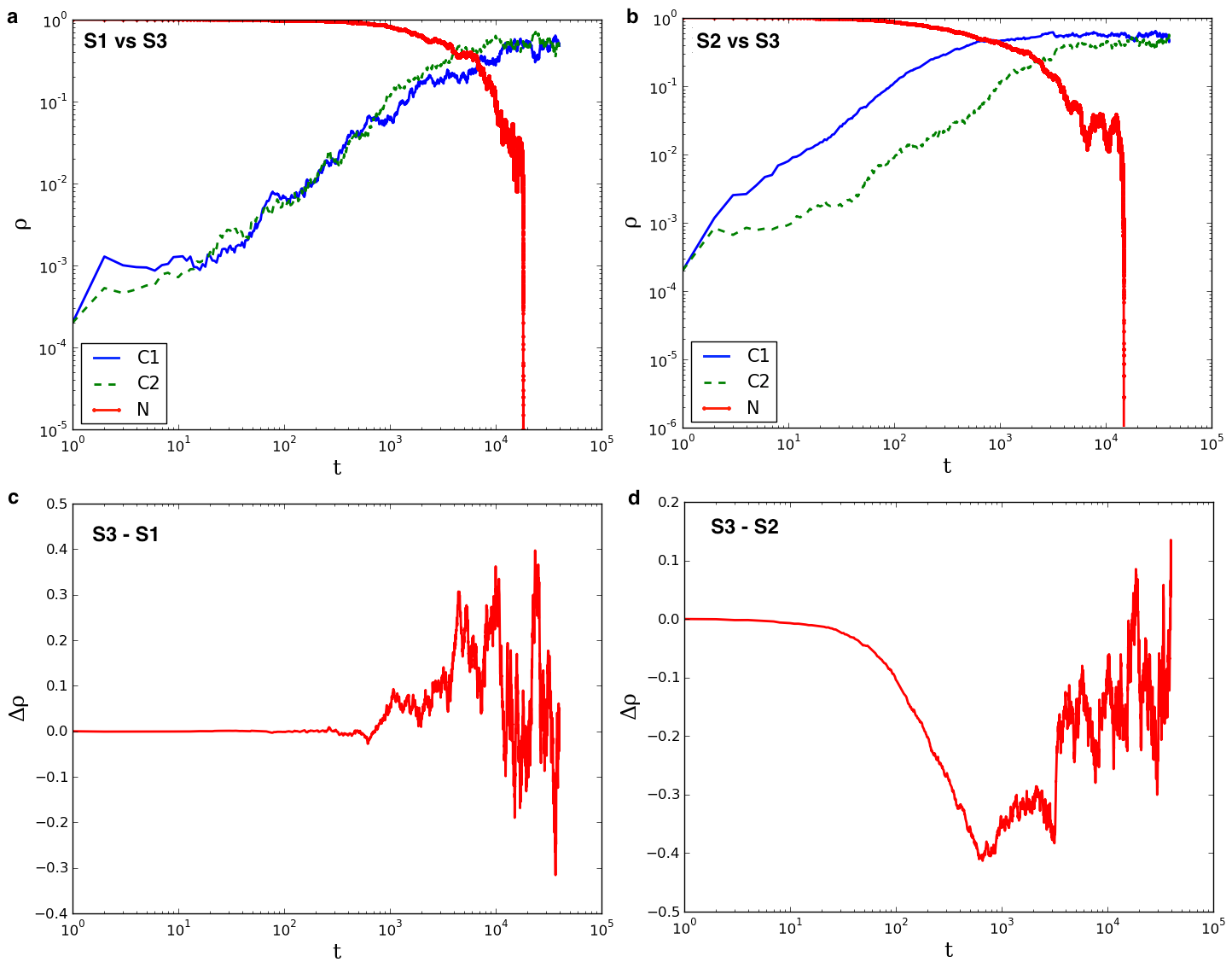}
\caption{\label{fig:figure_7} On the top: comparison among the density of agents in the state $0$ (labeled as $N$ in the legend, i.e., neutral agents), in the state $1$ (labeled as $C_1$ in the legend, i.e., Competitor $1$) and in the state $2$ (labeled as $C_2$ in the legend) over time, performed on scale-free networks. \textbf{a}) Competitor $1$ uses the strategy \textbf{S1} and competitor $2$ uses the strategy \textbf{S3}. \textbf{b}) Competitor $1$ uses the strategy \textbf{S2} and competitor $2$ uses the strategy \textbf{S3}. 
On the bottom: difference between densities of agents $\Delta\rho$ in the state $1$ and $2$ in scale-free networks. \textbf{c}) Results achieved when the competitors use the strategies \textbf{S3} and \textbf{S1}, respectively. \textbf{d}) Results achievedwhen the competitors use the strategies \textbf{S3} and \textbf{S1}, respectively. Results are averaged over $20$ different realizations.}
\end{figure*}

\begin{figure*}[!ht]
\centering
\includegraphics[width=6.0in]{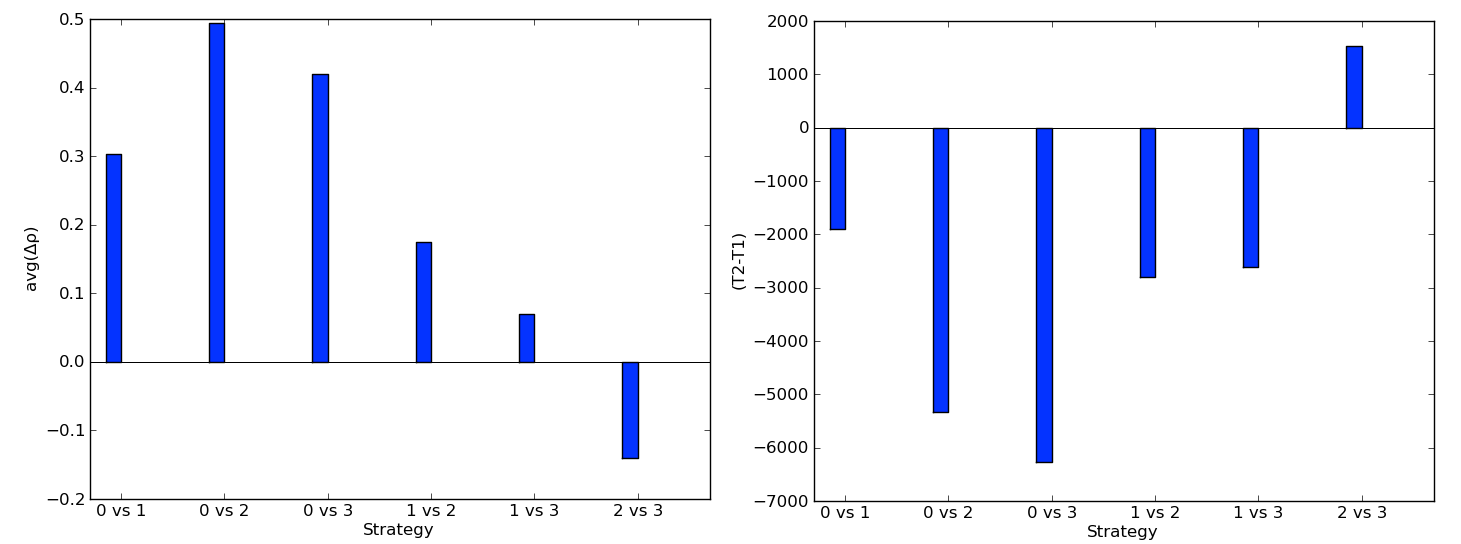}
\caption{\label{fig:figure_8} Results achieved in scale-free networks with $N=5\cdot10^4$ agents, varying the network strategy. On the left, $avg(\Delta\rho)$, i.e., average difference between densities of agents in the two states, $1$ and $2$. On the right, difference between points $T_2$ and $T_1$. Results are averaged over $20$ different realizations.}
\end{figure*}

In this section, we report results of the proposed model achieved by using scale-free networks with $N=5\cdot10^4$ agents. In so doing, we can perform a further evaluation on the effects of the hubs (i.e., nodes with a high degree) in these dynamics.
As indicated in Figure~\ref{fig:figure_7}, on a quality level, results are similar to those achieved in smaller scale-free networks --see Figures~\ref{fig:figure_2} and~\ref{fig:figure_4}.
We observe that increasing $N$ (i.e., the number of agents), the number of time steps to reduce neutral agents to zero increases. In particular, with $N=5\cdot10^4$, the density of neutral agents falls to zero after about $1.8\cdot10^4$ time steps, while with $N=10^4$ it takes about $5.5\cdot10^3$ time steps.
In Figure~\ref{fig:figure_8}, we show results related to parameters $avg(\Delta\rho)$ and $T_2 - T_1$. Also in these diagrams, we found results similar to those achieved in scale-free networks with $N=10^4$.
Therefore, we can state that also in the presence of more hubs in the agent network (considering the scale-free configuration), on a quality level, the outcomes of the proposed model are similar to those achieved in the main analysis.


\section*{References}

\begin{thebibliography}{10}
\bibitem{loreto01}
Castellano, C. and Fortunato, S. and Loreto, V.,
Statistical physics of social dynamics.
Rev. Mod. Phys. \textbf{81 - 2}, (2009) 591--646

\bibitem{galam01}
Martins, A.C.R., and Galam, S.,
Building up of individual inflexibility in opinion dynamics.
Phys. Rev. E \textbf{87-4}, (2013) 042807

\bibitem{galam02}
Galam, S.,
SOCIOPHYSICS: a review of Galam models.
International Journal of Modern Physics C \textbf{3}, (2008) 409--440

\bibitem{krapviski01}
Krapivsky, P. L. and Redner, S.,
Dynamics of Majority Rule in Two-State Interacting Spin Systems.
Phys. Rev. Lett. \textbf{90-23}, (2003) 238701

\bibitem{sznajd01}
Sznajd-Weron, K. and Sznajd, J.,
Dynamics of Majority Rule in Two-State Interacting Spin Systems.
International Journal of Modern Physics C \textbf{11-6}, (2000) 1157

\bibitem{holme01}
Holme, P. and Newman, M. E. J.,
Nonequilibrium phase transition in the coevolution of networks and opinions.
Phys. Rev. E \textbf{74-5}, (2006) 056108

\bibitem{bianconi01}
Halu, A. and Zhao, K. and Baronchelli, A. and Bianconi, G.,
Connect and win: The role of social networks in political elections.
Europhysics Letters \textbf{102 - 1}, (2013) 16002

\bibitem{miguel01}
San Miguel, M. and Eguiluz, V.M. and Toral, R.,
Binary and Multivariate Stochastic Models of Consensus Formation.
Computing in Science and Engineering \textbf{7 - 6}, (2005) 67 -- 73

\bibitem{barrat01}
Kozma, Balazs and Barrat, Alain,
Consensus formation on adaptive networks.
Phys. Rev. E \textbf{77-1}, (2008) 016102

\bibitem{redner01}
Sood, V. and Redner, S.,
Voter Model on Heterogeneous Graphs.
Phys. Rev. Lett. \textbf{94 - 17}, (2005) 178701

\bibitem{galam03}
Galam, S.:
From 2000 Bush–Gore to 2006 Italian elections: voting at fifty-fifty and the contrarian effect.
\emph{Quality \& Quantity} \textbf{41-4} 579-589 (2007)

\bibitem{serrano01}
Serrano, M.A., Klemm, K., Vazquez, F., Eguiluz, V.M., San Miguel, M.,
Conservation laws for voter-like models on random directed networks.
J. Stat. Mech. (2009) P10024

\bibitem{mobilia01}
Mobilia, M. and Redner, S.,
Majority versus minority dynamics: Phase transition in an interacting two-state spin system.
Phys. Rev. E \textbf{68-4}, (2003) 046106

\bibitem{vespignani01}
Castellano, C., Marsili, M., and Vespignani, A.,
Nonequilibrium Phase Transition in a Model for Social Influence.
Phys. Rev. Lett. \textbf{85-16}, (2003) 3536--3539

\bibitem{schweitzer01}
Schweitzer, F., and Behera, L.,
Nonlinear voter models: the transition from invasion to coexistence.
The European Physical Journal B \textbf{67-3}, (2009) 301--318

\bibitem{gross01}
Zschaler, Gerd and B\"ohme, Gesa A. and Sei\ss{}inger, Michael and Huepe, Cristi\'an and Gross, T.,
Early fragmentation in the adaptive voter model on directed networks.
Phys. Rev. E \textbf{85-4}, (2012) 046107

\bibitem{miguel02}
Fernandez-Gracia, J. and Suchecki, K. and Ramasco, J.J. and San Miguel, M. and Eguiluz, V.M.,
Is the Voter Model a model for voters?.
http://arxiv.org/abs/1309.1131, (2013)

\bibitem{petersen01}
Mobilia, M., Petersen, A., and Redner, S.,
On the role of zealotry in the voter model.
J. Stat. Mech. (2007) P08029

\bibitem{galam04}
Galam, S., Chopard, B.,  Masselot, A., and Droz,M.,
Competing species dynamics: Qualitative advantage versus geography.
The European Physical Journal B \textbf{4-4}, (1998) 529--531

\bibitem{clifford01}
Clifford, P. and Sudbury, A.,
A model for spatial conflict.
Biometrika \textbf{60-3}, (1973) 581--588

\bibitem{miguel03}
Castello, X. and Eguiluz, V.M. and San Miguel, M.,
Ordering dynamics with two non-excluding options: bilingualism in language competition.
New Journal of Physics \textbf{8}, (2006)

\bibitem{marvel01}
Marvel, Seth A. and Hong, Hyunsuk and Papush, Anna and Strogatz, Steven H.,
Encouraging Moderation: Clues from a Simple Model of Ideological Conflict.
Phys. Rev. Lett. \textbf{109-11}, (2012) 118702

\bibitem{gross02}
Gross, T. and Hiroki, S.,
Adaptive Networks: Theory, Models and Applications.
Springer Berlin Heidelberg, (2009)

\bibitem{gross03}
Gross, T. and Blausius, B.,
Adaptive coevolutionary networks: a review.
Journal of the Royal Society Interface \textbf{5-20}, (2008) 259--271

\bibitem{zschaler01}
Zschaler, G.,
Adaptive-network models of collective dynamics.
The European Physical Journal Special Topics \textbf{211-1}, (2012) 1--101

\bibitem{szolnoki01}
Szab\'o, Gy\"orgy and Szolnoki, Attila,
Three-state cyclic voter model extended with Potts energy.
Phys. Rev. E \textbf{65-3}, (2002) 036115

\bibitem{barabasi01}
Barabasi, A.L. and Albert, R.,
Emergence of scaling in random networks.
Science \textbf{286 - 5439}, (1999) 509--512

\bibitem{watts01}
Watts, D. J. and Strogatz, S. H.,
Collective dynamics of ``small-world'' networks.
Nature \textbf{393}, (1998) 440--442

\bibitem{hillygus01}
Hillygus, D.S.,
Encouraging Moderation: Clues from a Simple Model of Ideological Conflict.
Public Opinion Quarterly \textbf{75-5}, (2011) 962--981

\bibitem{merolla01}
Merolla, J.L. and Zechmeister, E.J.,
The Nature, Determinants, and Consequences of Chávez?s Charisma: Evidence From a Study of Venezuelan Public Opinion.
Comparative Political Studies \textbf{44-1}, (2011) 28--54

\bibitem{keyman01}
Muftuler-Bac, M. and Keyman, E.F.,
The Era of Dominant-Party Politics.
Journal of Democracy \textbf{23}, (2012) 85--99

\end{thebibliography}
\end{document}